\documentclass[preprint2]{aastex}
\usepackage{natbib}
\usepackage{bm}
\usepackage[dvips]{color}
\newif\ifAMStwofonts
\AMStwofontstrue 

\def\vPsi{{\bf \Psi}}
\def\vPhi{{\bf \Phi}}
\def\vmu{{\bm \mu}}
\def\vtheta{{\bm \theta}}
\def\vp{{\bf p}}

\def\tI{{\tilde I}}

\def\vv{\bf v}

\def\vB{{\bf B}}

\def\gsim{~\rlap{$>$}{\lower 1.0ex\hbox{$\sim$}}}

\def\simpropto{\lower.2ex\hbox{$\; \buildrel \propto \over \sim \;$}}
\def\ltsim{\lower.5ex\hbox{$\; \buildrel < \over \sim \;$}}
\def\gtsim{\lower.5ex\hbox{$\; \buildrel > \over \sim \;$}}
\def\ltsim{\lower.5ex\hbox{$\; \buildrel < \over \sim \;$}}
\def\gtsim{\lower.5ex\hbox{$\; \buildrel > \over \sim \;$}}

\def\kms{\mbox{km\,s$^{-1}$}}

\def\dd{\,{\rm d}}

\def\kms{\ {\rm km\,s^{-1}}}

\def\tl{ {\tilde l}}

\def\dd{{\rm d}}

\def\grad{\nabla}
\def\pmb#1{\setbox0=\hbox{#1}%
\kern-.025em\copy0\kern-\wd0
\kern.05em\copy0\kern-\wd0
\kern-.025em\raise.0433em\box0}

\def\vv{\pmb{$v$}}

\def\vs{\pmb{$s$}}

\def\vr{\pmb{$r$}}

\def\simlt{\lower.5ex\hbox{$\; \buildrel < \over \sim \;$}}
\def\simgt{\lower.5ex\hbox{$\; \buildrel > \over \sim \;$}}

\newcommand{\beq}{\begin{equation}}
\newcommand{\eeq}{\end{equation}}
\def\beqa{\begin{eqnarray}}
\def\eeqa{\end{eqnarray}}
\def\fixit#1{}

\def\dd{{\rm d}}

\def\muasyr{\mu as \, {\rm yr}^{-1} }

\shorttitle{Large scale flows from astrometry }
\shortauthors{Nusser, Branchini \& Davis}
\begin{document}
\title{Gaia: a Window to Large Scale Motions}
\author{Adi Nusser\altaffilmark{1}}
\affil{Physics Department and the Asher Space Science Institute-Technion, Haifa 32000, Israel}
\author{Enzo Branchini\altaffilmark{2}}
\affil{Department of Physics, Universit\`a Roma Tre, Via della Vasca Navale 84, 00146, Rome, Italy}
\affil{INFN Sezione di Roma 3, Via della Vasca Navale 84, 00146, Rome, Italy}
\affil{INAF, Osservatorio Astronomico di Brera, Milano, Italy}
\author{Marc Davis\altaffilmark{3}}
\affil{Departments of Astronomy \& Physics, University of California, Berkeley, CA. 94720}
\altaffiltext{1}{E-mail: adi@physics.technion.ac.il}
\altaffiltext{2}{E-mail: branchin@fis.uniroma3.it}
\altaffiltext{3}{E-mail: mdavis@berkeley.edu}


\begin{abstract}

Using redshifts as a proxy for galaxy distances, estimates of the 2D  transverse peculiar velocities 
of distant galaxies  could  be obtained from future measurements of 
proper motions. We provide the mathematical framework for analyzing 2D transverse motions and 
show  that they offer several advantages over traditional probes of large scale motions. They are  completely independent of 
any intrinsic relations between galaxy properties, hence they are  essentially free of selection biases. 
They are  free  from { homogeneous and } inhomogeneous Malmquist biases that typically plague distance indicator
catalogs.
They  provide  additional information  to traditional probes which 
yield line-of-sight peculiar velocities only.  Further, because of their 2D nature, fundamental 
questions regarding vorticity of large scale flows can be addressed. 
Gaia for example  is expected to   provide  proper motions of at least  bright galaxies with high central surface brightness,
making  proper motions a likely contender  traditional probes  
based on current and future  distance indicator measurements.

 \end{abstract}


\section{Introduction}
\label{sec:int}
 In the standard cosmological paradigm,  peculiar  motions (i.e. deviations from Hubble flow) of galaxies are the result of the process of gravitational instability with overdense regions attracting material, and underdense regions repelling material.  
 The coherence and amplitude of galaxy flows are a direct indication of the distribution of the dark matter, the cosmological background, { and the underlying theory of gravity} 
 Traditionally the peculiar velocity field is derived from observations of distance indicators such as the Tully-Fisher relation \citep{TFR}  between 
 luminosity and rotational velocity of galaxies. The observed flux and rotational velocity are then used to infer the distance from 
 the TF relation. The distance is  then subtracted from the redshift, $cz$, in order to obtain the line-of-sight component 
 of the peculiar velocity of a galaxy, with a typical $1\sigma$ error $\sim 0.2 \ cz $.
 
 Here we point out an alternative probe of the large scale velocity field by means of future likely measurements 
 of proper motions of galaxies. 
 As an example for such future measurements we consider 
 the Gaia\footnote{http://sci.esa.int/science-e/www/area/index.cfm?fareaid=26} space astrometric mission.
 Although the main aim of the mission is to study our Galaxy, Gaia will also be able to perform
 accurate astrometry for external galaxies, largely thanks to its excellent angular resolution,
 provided they are sufficiently distant \citep{perryman01,vaccat}.
As an example, the nuclei of M87 at N5121, both at $d \simeq 17.8 \ h_{70}^{-1} \  {\rm Mpc }$, 
will be detected with apparent magnitudes $V\sim 16$ and $V\sim 14.7$, respectively,  within
an aperture of 0.65 $arcsec$, approximately corresponding to Gaia's detection window
\citep{ferrarese94,carollo98,lauer07}.
With an expected  end-of mission  accuracy in the measurements of proper motions of $\sim (10 - 20) \  \muasyr$
at the $V$ magnitudes of these two galaxies, Gaia will be able to measure the
transverse displacements of these objects with an accuracy of (0.8-1.6) $10^{-4}  \ h_{70}^{-1} \  {\rm pc} $, 
corresponding to a transverse velocity $\sim 600 \kms$, which is a rather typical value,
comparable to that of the Local Group velocity with respect to the CMB.
Although Gaia's on board thresholding algorithm is optimized for stellar objects,
a large number of galaxies 
will have their stellar light emission concentrated in a compact region 
(either  a bulge or pseudo-bulge) of sub kpc in effective radius, sufficient to make them appear as detectable
point-like sources \citep[e.g.][]{korm77,allendriver06,ohama09,graham}.
\cite{robin12} estimates that Gaia will be able to detect $\sim 10^6$ galaxies.
In fact, high surface brightness substructures within extended objects might be detected 
as individual sources associated to the same galaxy, hence improving 
 the accuracy in the measurement of its peculiar motion, as we shall demonstrate in this paper.

Distances to galaxies are needed to derive their  transverse peculiar velocities  (in $\kms$) from the  proper motions.
High precision distances will become available for those star-forming galaxies that, once observed by Gaia, will
subsequently have their Cepheid distances determined (for example by JWST \footnote{http://www.jwst.nasa.gov/}). 
However, these will be available for a limited number of relatively nearby galaxies, whereas we are interested
in tracing the cosmic velocity field over large regions.
As a proxy for the distance we will use the 
galaxy's redshift,  which differs from the actual distance by the radial peculiar velocity.
The relative error in the transverse velocity as a result of this approximation is small and decreases with redshift. 
An object with  $V=15$ will  have an error in  transverse velocity of  $\sim 0.6 \ cz$. 
This is significantly larger than the uncertainty in  line-of-sight peculiar velocities from 
distance indicators.
However, we will show that the large number of galaxies expected to be observed with Gaia 
will beat the increased scatter, possibly making Gaia's proper motions an excellent probe of the large scale flows. 
This  probe of large scale flows   is completely independent of 
 any assumption on the intrinsic relations of galaxies. Further, 
 the 2D transverse motions are orthogonal (in information content as well as in geometry) to  
 standard line-of-sight peculiar velocities.

The outline of the paper is as follows. In \S\ref{sec:meth} 
we present the general set up and describe theoretical tools for analyzing future transverse velocity data. 
We present, in \S\ref{sec:err}, a rough estimate of the expected error in the  transverse velocity obtained by smoothing individual 
velocities. Expected errors on astrometry for Gaia's galaxies are discussed in \S\ref{sec:ext1} and a more general discussion
 on astrometry of extended objects is given in \S\ref{sec:ext2}. In the concluding section  \S\ref{sec:cnl}, we present  a general assessment of the  transverse velocity data in comparison to other  probes of 
large scale motions. We also discuss possible sources for redshifts
 of the population of  galaxies expected to be observed by Gaia. 
 
Unless otherwise specified,  magnitudes observed by Gaia will 
refer to an aperture photometry 0.65 $arcsec$. They are given in the $G$ band (350-1000 nm). 
Transformation from the more familiar $V$ and $I_c$ bands are performed using constant colors $V-G=0.27$ 
and $V-I_c =1$ for all galaxies \citep{fukug,jordi10}.
We also assume that Gaia will identify all sources with $G<20$ within  0.65 $arcsec$ with 
 100 \% completeness. Finally,  we use $H_0=70\kms {\rm Mpc}^{-1}$ to set the distance scale 
 and use $h_{70}=H_0/70$ to parametrize uncertainties.

\section{Methodology}
\label{sec:meth}
We will assume an all-sky catalog of redshifts and proper motions. 
 We denote the physical peculiar velocity by $\vv$ and the real space comoving coordinate by 
 $\vr$, both expressed in $\kms$. 
 Further,  $v_\parallel=\vv \cdot \hat {\vr}$ and $\vv_\perp=\vv-v_\parallel \hat {\vr}$
  are, respectively,  the components of $\vv$ parallel and perpendicular to the line-of-sight, where $\hat {\vr}$
   is a unit vector in the line-of-sight direction.
 We restrict the analysis to $ cz \ltsim 15,000\kms$ and neglect cosmological geometric effects,
  so that the redshift coordinate is $ \vs=\vr+v_\parallel \hat {\vr}$. Note $\hat {\vs}=\hat {\vr}$ and  $cz=r+v_\parallel=\vs \cdot \hat {\vr}= s$.  
Proper motions { transverse to the line-of-sight} will be denoted by $\vmu$. 
The transverse 2D space velocity is of a galaxy at real space distance $r$ is 
 \begin{eqnarray}
 \vv_\perp&=& r \vmu\\
 \nonumber &=&677.22 \frac{\vmu}{1\muasyr} \frac{r}{10^4 \kms} h_{70}  \; ,
 \end{eqnarray}
  which corresponds a  transverse peculiar velocity of $474\kms$ for  $1 \ \muasyr$ at $d=100$ Mpc. 
  
  However,  the true distances, $r$, are unknown, and, therefore, we make the approximation 
\begin{equation}
\label{eq:perp}
\vv_\perp={s}\vmu\; . 
\end{equation}
This  introduces a relative error $v_\parallel/s$ in the determination of $\vv_\perp$ where $<v_\parallel^2>^{1/2} \sim  200-300\kms$ \citep{DN10}.
Hence the error is negligible as we go to $ s\gtsim 2,000\kms$.  The error is also random since 
$  <\vv_\perp v_\parallel> =0$.

Therefore, the estimated velocity field will be given as a function 
of the redshift space coordinate. To linear order,  velocity fields expressed in real and redshift 
spaces are equivalent. In the quasilinear regime, dynamical relations 
can be derived for the velocity field in redshift space \citep{ND94}, thanks to the interesting property that an irrotational (or potential) flow in real space remain irrotational also in redshift space \citep{chodnuss}.
\subsection{From 2D transverse velocities to 3D flows}

Here we offer basic  expressions for the derivation of the full peculiar velocity field  $\vv(\vs)$ from the 
smoothed 2D transverse velocity field, $\vv_\perp(\vs)$. 
Assuming a  potential flow $\vv(\vs)=-\grad \Phi(\vs)$
and expanding  the angular dependence of $\Phi$ in spherical harmonics,
 $\Phi(\vs)=\sum_{l m}\Phi_{lm}(s)Y_{lm}(\hat {\vs})$, gives \citep{ARFKEN}
\begin{equation}
v_\parallel=-\sum_{l m}\frac{\dd \Phi_{lm}}{\dd s}  Y_{lm} 
\end{equation}
\begin{equation}
\vv_\perp=-\sum_{lm} \frac{\Phi_{lm}}{s} \vPsi_{lm} \; ,
\end{equation}
where $\vPsi_{lm}=r \grad Y_{lm}$ is the vector spherical harmonic.
Thanks to the orthogonality conditions 
$\int \dd \Omega \vPsi_{lm} \cdot \vPsi_{l'm'}=l(l+1)\delta^K_{ll'}\delta^K_{mm'}$
the potential coefficients can be recovered by 
\begin{equation}
\label{eq:phfromv}
\Phi_{lm}(s)=\frac{-1}{l(l+1)}\int\dd \Omega \vv_\perp(\vs) \cdot \vPsi_{lm}(\hat {\vs}) \; ,
\end{equation}
for $l>0$. This means that $\Phi(\vs)$ can be recovered from the 
$\vv_\perp$ up-to a monopole term which corresponds to  a purely radial flow with zero transverse motions. That is not a serious drawback since 
the monopole term can always be removed from the predictions of any model to be compared with the 
data. 

\subsection{Testing the potential flow ansatz }
Initial conditions in the early Universe might have been somewhat chaotic, so that the original peculiar velocity field  was uncorrelated with the mass distribution, or even contained vorticity \citep[e.g.][]{christvor}. 
At late time, a cosmological velocity field should have a negligible rotational component, $\vv^{\rm rot}$  on large scale, away from 
orbit mixing regions. The reason is that any 
   circulation, $\Gamma=\oint \vv^{\rm rot}\cdot \dd \vs$, is  conserved by Kelvin's theorem.
 Hence, any 
rotational component will decaying as   $1/a$, where $a$ is the scale factor. In contrast, 
the irrotational component of the peculiar velocity will have a growing $v\sim \sqrt{a}$. 
Therefore,  on large scales, away from collapsed objects, the irrotational component is expected to be 
negligible. The absence of any significant  large scale vorticity is, therefore, a strong prediction 
of the standard cosmological paradigm. 
To  assess  this prediction,  the observed transverse motions can be used to  constrain the amplitude of the irrotational 
component. This can be done by writing the transverse component of $ \vv^{\rm rot} $ 
as \citep{ARFKEN}
\begin{equation}
\vv^{\rm rot}_\perp=\sum_{lm} V^{\rm rot}_{lm}\vPhi_{lm}\; ,
\end{equation}
where $\vPhi_{lm}=\vs \times \grad Y_{lm}$ belong to  another class  of vector spherical harmonics
that satisfy the same orthogonality conditions as $\vPsi$. Hence,
$V^{\rm rot}_{lm} $ is equal to the r.h.s of Eq.~ \ref{eq:phfromv} but with   $\vPhi_{lm}$
instead of $\vPsi_{lm}$.
Further, 
 $\int \dd \Omega \vPhi_{lm} \cdot \vPsi_{l'm'}=0$, hence the recovery of the rotational mode is formally independent 
 of the potential flow mode.

\section{The expected errors } 
\label{sec:err}

We provide estimates of the expected random errors in the smoothed transverse velocity field,  $\vv_\perp(\vs)$, 
as a function of distance from the observer.\footnote{ In generating the smoothed $\vv_\perp(\vs)$ care must be employed 
since the transverse directions of galaxies in different sight-lines within a filtering window do not 
point in the same direction.
This difficulty could be overcome by 
tensor window smoothing \`a la  POTENT 
\citep{DBF}. However,  we will not be concerned with these fine details at this stage. }

\begin{figure*} 
\centering
\epsscale{1.5}
\plotone{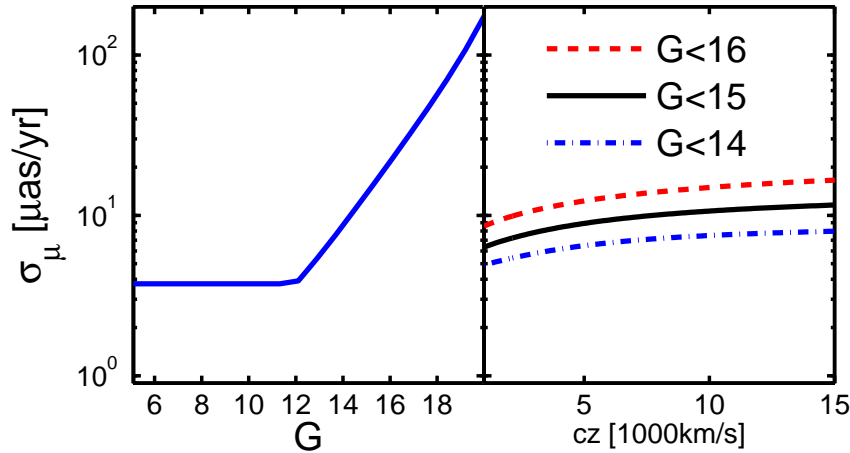}
\vspace{0pt}
\caption{Expected error  in Gaia's proper motion measurements.
{\it Left}:   $1\sigma$ error of an object as a function of its G magnitude.
  {\it Right:} mean  $1\sigma$ error  of objects as a function of distant  for three 
  values of the G magnitude cut. }
\label{fig:sigu}
\end{figure*}

The expected $1\sigma $ error, $\sigma_{\mu }(m)$, in the measurement of an object's proper motion 
depends on its $G$ magnitude and, to a lesser extent, on its color \citep{debru}.
Hereafter, we will use   $\sigma_{\mu } $ as a function of $G$, according to  the expression referenced in \cite{debru}. 
This $\sigma_\mu(m)$  is plotted in the left panel of Fig.~\ref{fig:sigu}. 
Other possible error sources are photometric jitter from SNe, AGN and,  for the latter sources, 
the presence of  jets with large proper motions.
We assume that spectrophotometry available for all objects detected by Gaia will significantly reduce the 
impact of these error sources which, therefore, will be neglected in the error budget.

The $rms$ accuracy in the galaxy proper motion at redshift $s$ can be obtained by summing over 
all galaxy magnitudes
\begin{equation}
\label{eq:scz}
\langle \sigma_m^{-2}(s)\rangle=\int^{m_{lim}}_{-\infty} n(m,s) \sigma_\mu^{-2}(m) \d dm \;.
\end{equation}
We assume that the surface brightness [SB] profile of distant galaxies is 
sufficiently peaked to guarantee that a large fraction of the galaxy luminosity is
within Gaia's detection window. The validity of this hypothesis will be discussed in
Section~\ref{sec:ext1}. In this case the number density of galaxies $ n (m,s)$, is simply related to the galaxy luminosity function 
 $N(M)$:   $n(m,s) \dd m=N(M)\dd M$,  where  $M=m-5{\rm log}_{10}(s)-15$ is the absolute magnitude.
For  Gaia's G band, we approximate $N(M)$ by the Schechter form of  the $V$ band luminosity function
with parameters given by \citep{brown01}.
Other choices for the Schechter parameters  of the $V$ band luminosity function
 \citep{marchesL}  do  not change the results significantly at the magnitude limits considered here.
The results for the average error are shown in the right panel for 
three magnitude cuts. 
The flatness of the curves for all magnitude, is 
a reflection of the fact that number of galaxies increases strongly with magnitude. 

Given individual measurements $\vv_{\perp i}=\vv_{\perp}(\vs_i)$, we  write the smoothed velocity as 
\begin{equation}
\label{eq:smoo}
\vv_\perp(\vs)=\frac{\sum_i {\vv_{\perp i}}{\sigma_{\perp i}^{-2} } W(\vs,\vs_i) }{\sum_i \sigma_{\perp i}^{-2 } W(\vs,\vs_i)}
\end{equation}
where the summation is over all galaxies,  $\sigma_{\perp i}=s\sigma_{\mu i}$
 and $W$ is a smoothing window function. 

The 1$\sigma$ errors on $\vv_\perp(\vs)$ is given by
\begin{equation}
\sigma^2_\perp(\vs)=\frac{\sum_i {\sigma_{\perp i}^{-2} } W^2(\vs,\vs_i) }{\left[\sum_i \sigma_{\perp i}^{-2 } W(\vs,\vs_i)\right]^2}\; . 
\end{equation}
The summation over galaxies can be transformed into a volume  integration with  the same argument but 
multiplied by the number density of galaxies. 
Doing so for a uniform distribution and assuming a  Gaussian window, $W$, of width $R_{G}$, we get

\begin{equation}
\label{eq:sigerr}
\sigma^2_\perp(s)=\frac{ s^2}{8\pi^{3/2} R_G^3  }
\frac{1}{\langle \sigma_m^{-2}(s)\rangle}\; ,
\end{equation}

\begin{figure} 
\centering
\epsscale{1}
\plotone{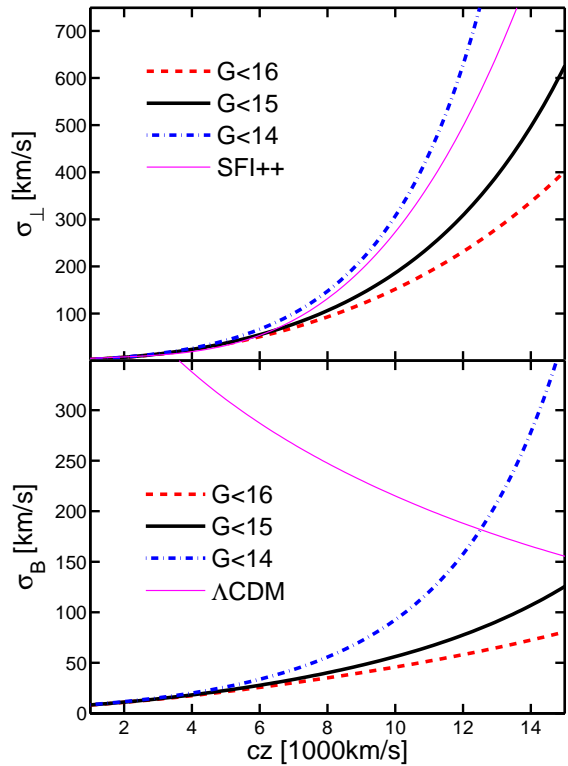}
\vspace{0pt}
\caption{Expected errors (1$\sigma$) on two quantities computed from the Gaia 
astrometric galaxy data.
{\it Top}: Errors in the 2D transverse peculiar velocity field  
obtained by filtering the data with  a gaussian window of width  $R_G=1500\kms$. 
For comparison, the thin solid magenta line is the error  in the SFI++ 
line-of-sight peculiar velocities smoothed with the same window. 
Errors scale like $R_G^{3/2}$.
{\it Bottom}: Errors in the bulk (dipole) motion of  spherical shells  of thickness  $\Delta cz=3,000\kms$.
Errors scale like $(\Delta cz)^{1/2}$. For reference, predictions from the WMAP7 $\Lambda$CDM
for the dipole on shells are also plotted. 
In both panels, dash-dotted, solid and dotted curves correspond to G=14, 15 \& 16 magnitude cuts, as indicated in the figure.   }
\label{fig:sigv}
\end{figure}

We have assumed the distant observer limit so that $|\vs_i-\vs|\ll s$. 
For a Top-Hat window of the same width we get the same expression but 
with  $4\pi/3$ as the numerical factor in the denominator.
Substituting  $\sigma_\mu(m)$ (see left panel Fig.~\ref{fig:sigu}) in Eq.~\ref{eq:sigerr}, we compute the expected 
error, $\sigma_\perp$,  in the smoothed $ \vv_\perp$,  for a gaussian smoothing  with $R_G=1,500\kms$.
The top panel in Fig.~\ref{fig:sigv} shows curves of $\sigma_\perp$ as a function of distance 
 for three magnitude cuts.

 For comparison the figure also plots the error in the filtered line-of-sight peculiar velocities in the SFI++ catalog of TF measurements of $\sim 4,000$ galaxies \citep{mas06}. 
There is a significance decrease in $\sigma_\perp$ as the magnitude is increased from $G=14$ to 15,
but the improvement is not as dramatic when fainter galaxies with $15<G<16$ are included. 
The reason is the rapid deterioration in $\sigma_\mu$ at $G=16$ which is not compensated by 
the added number of fainter galaxies.  At redshifts   $s\gtsim 6,000\kms$ and for  $G<15$, peculiar velocities from Gaia's proper motions are expected to   fair much better than the SFI++ catalog \citep{mas06}. 
Another important quantity which can be computed from transverse velocities is 
the dipole motion (i.e. bulk flow) of spherical shells of a given thickness. This 
motion is described by a constant term $\vB$ and gives rise to a transverse 
velocity field of the form $\vv_{\perp B}=\vB -\hat {\vr} (\vB \cdot \hat {\vr})$.
The dipole term $\vB$  can be found by least squares fitting of $\vv_{\perp B}$ to the 
observed velocities  $\vv_{\perp i}$.
The expected error in $\vB$ as a function of distance of the shell, is plotted in the bottom panel in Fig.~\ref{fig:sigv}.
Predictions for 3 magnitude cuts are plotted for 
spherical shells of $3,000\kms$ in thickness.
For comparison we also plot the 
WMAP7 $\Lambda$CDM model \citep{wmap7} predictions for the amplitude of the velocity dipole on spherical shells. 
It is encouraging that the predicted amplitude  is  larger than the expected error out to relatively large distances.

\section{Astrometry with Gaia's galaxies} 
\label{sec:ext1}

Here we provide a rough argument demonstrating the possibility of high precision 
astrometry with Galaxies observed by Gaia. 
To do this we use Gaia condition for astrometric measurements
($G<20$ within an aperture of $\sim 0.65 \ arcsec$) to define an analogous  threshold 
based on SB. The mean SB of a $G=20$ object within Gaia detection window is 
 $ \mu_G \sim20 \  {\rm mag} \ arcsec^{-2}$. Here we shall make a more conservative choice and 
 assume that only sources with  $\mu_G < 18.5 \  {\rm mag} arcsec^{-2}$ will be used for astrometric purposes.
A survey of the literature shows that this condition is 
satisfied for the central region of a significant fraction of galaxies
 \citep[e.g.][]{korm77,allendriver06,ohama09,balcells,sbsmith,graham,ferrarese94,carollo98,lauer07}.  
For example, this can be seen in figure 3 in \cite{ohama09} showing a scatter 
plot of the B band effective SB versus half light radius 
for various galaxy types\footnote{For old stellar populations, $B\sim V+1$ \citep{fukug}, and since $G=V+0.27$, the astrometric condition  $G\ltsim 18.5$ translates to $B\ltsim 19.7$.}.
More importantly, we have visually inspected the observed $V$-band SB
profiles of 200 out of $\sim 600$ galaxies  in the Carnegie-Irvine Galaxy Survey  \citep[CGS; ][]{HOCGS,LICGS}.  
Most of these galaxies are nearby  (median  distance of $\sim 25 \ h_{70}^{-1} \ {\rm Mpc}$) and with mean $B$-band absolute 
{\it total} magnitude of  $-20.2 $, close to $M_*$. We  identified galaxies reaching  central SB of $18.5 \ \rm mag/arcs^2$
and tabulated the corresponding radii (in $arcsec$).  Since we did not have access to the actual data 
the minimal radius we could determine using a ruler  is $1-2 \ arcsec$. 
About 70\% of the galaxies we inspected were brighter than
$18.5 \rm mag/arcs^2$ allowing them to be detected by Gaia. 
Since SB is a distance-independent quantity we can use this threshold
to compute the maximum distance at which a galaxy would be detected in a single resolution element of Gaia. 
We find that the majority of early and late  type galaxies could be detected as point sources  at $G=20$
if, respectively,  placed at $\gtsim 500 \ h_{70}^{-1} \ \rm Mpc$ and $\gtsim 250 \ h_{70}^{-1} \ \rm Mpc$.
Overall, it looks like the overwhelming majority
of early type galaxies and more than 50 \% of late types will have peculiar motions measured by Gaia
with errors in transverse velocities given in the top panel in Fig.~\ref{fig:sigv}.
In addiction, a significant fraction of their emitted light will be within Gaia's detection window,
which justify the simple relation between galaxy number density and luminosity function that we have
adopted in Section~\ref{sec:err}.
AGN will be easily detected by Gaia as bright, pointlike sources and possibly mistaken by galaxies.
However, their contamination to a relatively local sample of objects with measure redshift,
like the one we consider here,  should be negligible.

In fact, since we are interested in studying the velocity field of the local
($\ltsim 100 \ h_{70}^{-1}\  \rm Mpc$) Universe the situation is likely to be even more favorable.
Within this distance the typical galaxy will be resolved in high SB substructures 
that, if brighter than $G=20$, can be detected as individual sources and analyzed as a group.
Example of multiple high-SB sources are star forming regions, globular clusters and
bulges with steep SB profiles that are more extended than Gaia's window
(for example, the SB-profile of M87 drops below $18.5 \rm mag/arcs^2$ at $\sim 700 \ h_{70}^{-1} \  {\rm pc} $
from the center. 
If placed at  $\sim 50\  h_{70}^{-1} \ {\rm Mpc} $ it will be detected as $\sim 10$ individual sources by Gaia).
Detecting multiple sourcs from the same objects significantly improves the astrometric precision, as we shall show 
in the next Section.

\section{Astrometry with extended objects} 
\label{sec:ext2}

The possibility of placing multiple constraints on the same objects
allows, in principle, to improve the astrometric accuracy.
We discuss this possibility in a general context and with a formalism that contemplate 
both the possibility of performing resolved photometry with high resolution instruments
like HST\footnote{http://www.stsci.edu/hst/}, JWST, LSST or Pan-STARRS \citep{LSST, panstarrs} , 
and that of splitting an extended source in individual sources, like in the  case of Gaia.

Suppose for simplicity we observe a galaxy at two different epochs, $t_1$
and $t_2$. Let us define  $I_i(\vtheta_i)$ the SB of the object at the epoch  $t_i$
measured at  the angular position of a pixel $\vtheta_i$. In the case of traditional
photometry  $I_i(\vtheta_i)$ represents  the SB profile of the object at $\vtheta_i$
whereas in the case of Gaia it represent the magnitude of the SB-substructure
measured within the detection window.
 In principle the astrometric shift, $\vp$, could be determined by minimizing, with respect to $\vp$, 
$\chi^2=\sum_i[I_1(\vtheta_i)-I_2(\vtheta'_i)]^2/\sigma_i^2$ where the summation is over all pixels, $ \vtheta'=\vtheta-\vp$ and  $\sigma_{Ii}$  here is the  $1\sigma $ error  in the measurement of the SB (since $\vp$ is small we assume that $\sigma_{Ii}$ 
in pixel $i$ is the same for both images).  We have assumed that $I_1$ and $ I_2$ differ only by a linear displacement. 
In principle one should take into account changes in the internal structure of the object. Those, however, 
will have little effect compared to the overall observational accuracy. Since we will eventually be interested 
in the mean coherent displacement of an ensemble of many galaxies, incoherent changes in the 
internal structure of galaxies will be insignificant. 

This procedure of minimizing the image differences exploits  all information contained in both images but it requires a possibly non-trivial 
interpolation of $\vtheta'$  on the observed pixel positions at $\vtheta$. 
Therefore, we present here an alternative technique which alleviates this 
problem and clarifies additional matter related to the astrometric expected precision for extended objects. 
Suppose that the actual galaxy image at time $t_1$,  is described by 
$I_s(\vtheta)=\sum_\alpha c_\alpha \tI^\alpha(\vtheta)$ where $\tI^\alpha$ are basis functions which may chosen to be orthonormal. 
Any choice (e.g. Fourier modes, wavelets) for $\tl$ would do for our purposes here. 
The underlying image at $t_2$ is therefore   $I_s(\vtheta+\vp)$.  The modeling in terms of the  basis functions $\tI^\alpha$ should account that the signal is modulated by the PSF. while photometric noise is just white noise.  
The expansion coefficients $c_\alpha$ and the displacement $\vp$ are 
determined by minimizing 
\begin{eqnarray}
\label{eq:cht}
\chi^2&=&\sum_i \sigma_i^{-2} \left[  I_{1i}-\sum_\alpha c_\alpha \tI^\alpha_i \right]^2 \\
&+&\sum_i \sigma_{Ii}^{-2} \left[  I_{2i}-\sum_\alpha c_\alpha \left(\tI^\alpha_i +\frac{\partial \tI^\alpha_i}{\partial \vtheta}\vp\right)\right]^2\nonumber \; .
\end{eqnarray}

More generally, images are taken at many different epochs (about 70 epochs in the case of Gaia). Therefore, 
it is more appropriate to write $\vp=\vmu t_k$ and to minimize the total $\chi^2$ with respect to $\vmu$. Since the 
generalization is straightforward, for brevity of notation we adhere to the simple situation described by Eq.~\ref{eq:cht}. We note that several variations of this procedure could be adopted. For example,  as an alternative to minimizing $\chi^2$ in Eq.~\ref{eq:cht}, we could use basis functions defined 
in terms of  $\vtheta-\vtheta_c$ where $\vtheta_c$ is an assumed position comoving with a given point on the galaxy (e.g. the centroid in the case of a spherical object).  We then could minimize the counterpart of the  first term 
in Eq.~\ref{eq:cht} with respect to $c_\alpha$ and $\vtheta_c$ to get $\vtheta_c$ at epochs $t_1$ and $t_2$.
Using the same model for the images at the two epochs,  the difference between $\vtheta_c$ would then be the displacement $\vp$.
This will   yield identical results to  minimization of Eq.~\ref{eq:cht} of our choice of $\tI^\alpha$ 
given as functions of $\vtheta$ rather $\vtheta-\vtheta_c$.  The covariance matrix of the error in the estimated parameters $c_\alpha$ and $\vp$ 
is given by the inverse of the hessian  of $\chi^2$ formed from $H_{cc}=\partial^2\chi^2/\partial c_\alpha\partial c_\beta$, 
$H_{cp}=\partial^2\chi^2/\partial c_\alpha \partial \vp$, and  $H_{pp}=\partial^2\chi^2/\partial  \vp \partial \vp$.
It is easy to show that $|H_{cp}|\ll  |H_{pp}|$, 
implying that the error on $\vp$ is 
$H_{pp}^{-1}$, i.e. almost  independent of how well $c_\alpha$ are recovered. 
Considering a one dimensional displacement we get  an error 
of $\sigma_p^2\propto 1/(\sum_i (dI_s/d\theta)^2/\sigma_i^2)$. Assuming the objects' SB dominates 
the sky background so that $\sigma_{Ii}^2\propto I_s$, we get 
\begin{equation}
\sigma_p^2\propto \frac{1}{f <(dI_s/d\theta)^2/I_s^2>}\; ,
\end{equation}
where $f$ is the observed total  flux of the object. 
Note that the averaged quantity $<(dI_s/d\theta)^2/I_s^2>$ is independent of the amplitude of 
of $I_s$. Hence $\sigma_p$   depends on the total observed flux and variance in 
logarithmic derivative of the SB. The actual value of the SB is irrelevant as long as it satisfies the 
detection criteria. 
The larger the fluctuations/irregularities  in the stellar light, the more accurate is the astrometry. 
These irregularities may arise from different physical conditions in galaxies, e.g. spiral arms, young stellar associations, 
gravitational clumping  of stars, caustics  resulting from recent merging, and  patchy intrinsic dust obscuration. 
In the case of Gaia, they will be seen as individual sources with a S/N ratio $\gtsim 5$ at 
$1 \ arcsec^2$ at SB of $20 \ {\rm mag} \  arcsec^{-2}$ \citep{vaccat}.

 For  galaxies at $\ltsim 100 \ h_{70}^{-1} \  \rm Mpc$, SB fluctuations due to Poisson fluctuations in the finite number of (mainly hot luminous) stars per pixel \citep{sbf99,biscardi08} may contribute as additional source 
 of irregularities.  
 It is interesting that a sufficiently patchy extended object  brighter than $G\approx 12$ may yield 
 more accurate astrometry of a point source with the same luminosity. The reason is   
 the noise floor for point sources with $G\ltsim 12$.  
 An extended object of the same luminosity but made of numerous patches each with 
 $G>12$ could therefore yield higher precision than the point source.  We conclude that  astrometry of 
extended objects could well be comparable to those of point sources.

\section{Discussion}
\label{sec:cnl} 
The number of galaxies expected  to be observed by 
Gaia is likely to exceed standard distance indicator data by two orders of magnitude
 \citep{mas06,springsixdf,mastwomtf,tullycf}.
 Pan-STARRS and LSST \citep{panstarrs,LSST}  will yield 
 a factor of $100$ 
 less accurate proper motions ($\sim{\rm m}as \ yr^{-1}$) than Gaia,  but they will have
substantially more galaxies and therefore will also be useful for large scale motions.
Despite the larger object-by-object error, the large number of galaxies in catalogs of proper motions make them  a serious contender to traditional probes of the peculiar velocity 
field. The method has several advantages. Firstly, it is completely independent of 
any assumed intrinsic relations of galaxies and, hence, it does not suffer from the usual concerns 
related to these relations, e.g. linearity, selection biases and  dependence on environment. 
Secondly, it yields the 2D transverse velocity component and hence it offers  completely 
orthogonal information to standard probes which yield the line-sight-component. 
Thirdly, it is free from homogeneous and inhomogeneous Malmquis biases  \citep{lyn88}.

The usefulness of the method for probing the 3D velocity field on scales of a few 10s of 
Mpc is limited to $s \ltsim 10,000\kms$.
However, large scale  moments of the velocity field can be assessed at much larger distances. 
In particular the error of the dipole on (i.e. bulk flow of) spherical shells
can be estimated with $\sim 100-200\kms$ error at $s \sim 15,000 \kms$.
At larger redshifts, neither this method nor traditional ones are 
comparable to  the constraints on the dipole from galaxy luminosities in future galaxy redshift 
surveys\citep{NBDL}. 
At lower distances   ($ < 2,000 \kms$)  the 
transverse motions of galaxies could play an important role at providing 
 new constraints on  the motion of Local Group of galaxies.
 
Gaia will provide spectroscopic information of unresolved galaxies 
 \citep{tsalmantza}, especially  those with a high SB nucleus that will be preferentially detected.
However, the  inferred redshifts may not be sufficiently accurate or available 
for all unresolved galaxies with astrometric data
 \citep[e.g.][]{robin12}. The  Two Mass Redshift Survey \citep[2MRS;][]{huch2011} offers
 redshifts of $\sim 4\times 10^4$ galaxies down to $K_s=11.75$. This is the deepest all-sky redshift 
 catalog currently available.
It was originally planned to reach $Ks=12.2$ mag and to include $\sim 10^5$ galaxies. 
This is similar to the expected number of galaxies detected by Gaia brighter than $V=15.27$ (i.e. $G=15$)
 Further,   since $ Ks \approx K$ \citep{carpen01}
and $V-K\sim 2.7$ for most galaxies \citep{aa78}, we conclude that the $Ks = 12.2$ 2MRS has the same
objects' number density as the Gaia galaxies observed to $G = 15$, and undoubtedly it is largely the same sample. 
This is particularly interesting as Gaia's astrometric 
accuracy deteriorate rapidly at fainter objects.
 However,  it is unclear if  2MRS  will be continued to $K_s=12.2$  in the very near future (Macri, private communication). 
 For the purpose of the analysis presented here one could use a catalog of photometric redshifts based on the 2MASS galaxy catalog (Skrutskie et al. 2006), containing almost 1 million sources with $K_s<13.5$ mag. Its current form (2MASS XSCz, \cite{jarrett04}) offers distance errors as large as ~20-25\%,   which will improve in the coming years using the data from other galaxy catalogs for the photo-z estimation (Bilicki, private communication).
 
We have restricted the error analysis here to $G\sim 15$ since redshifts 
will probably not be available for all fainter  galaxies. 
However, data at fainter magnitudes can well be exploited
by computing the  dipole as a function of an effective depth corresponding to a certain magnitude range. 
This can then be compared with model predictions for an equivalent quantity  \citep{bilicki11}.

\section{Acknowledgments}
We thank Maciej Bilicki and Gary Mamon for useful comments. 
 This work was supported by THE ISRAEL SCIENCE FOUNDATION (grant No.203/09), the German-Israeli Foundation for 
Research and Development and  the Asher Space Research
Institute.
MD acknowledges the support provided by the NSF grant  AST-0807630. 
EB  acknowledges the support provided by the Agenzia Spaziale Italiana (ASI, contract N.I/058/08/0)
EB thanks the Technion Physics Department
for the kind hospitality. AN is grateful for the hospitality of  the Physics Department, Universit\`a Roma Tre.
EB thanks Michele Bellazzini and Paola Parma for suggestions and discussions.

\bibliography{gaia.bib}
 
\end{document}